\documentclass[pdftex,twocolumn,epjc3]{svjour3}          

\RequirePackage[T1]{fontenc}
\usepackage{placeins}
\usepackage{amsmath}

\smartqed  

\RequirePackage{graphicx}
\RequirePackage{mathptmx}      
\RequirePackage{flushend}
\RequirePackage[numbers,sort&compress]{natbib}
\RequirePackage[colorlinks,citecolor=blue,urlcolor=blue,linkcolor=blue]{hyperref}

\journalname{}

\begin{document}

\title{Lightweight Deep Learning Framework for Accurate Particle Flow Energy Reconstruction}

\author{Yu Wang\thanksref{e1,a,addr1}
        \and
        Yangguang Zhang\thanksref{e2,a,addr2}
        \and
        Shengxiang Lin\thanksref{e3,a,addr3}
        \and
        Xingyi Zhang\thanksref{e4,a,addr4}
        \and
        Han Zhang\thanksref{e5,a,addr5}
}
\thankstext{e1}{e-mail: u202342320@xs.ustb.edu.cn}
\thankstext{e2}{Corresponding author e-mail: 120221010614@ncepu.edu.cn}
\thankstext{e3}{e-mail: linshengxiang@stu.xjtu.edu.cn}
\thankstext{e4}{Corresponding author e-mail: zkevxinyg@sjtu.edu.cn}
\thankstext{e5}{e-mail: 2321010206@hhu.edu.cn}
\thankstext{a}{Yu Wang, Yangguang Zhang, Shengxiang Lin, and Xingyi Zhang  contributed equally to this work and should be considered co-first authors.}

\institute{
          School of Automation and Electrical Engineering, University of Science and Technology Beijing, Beijing, 100083, China\label{addr1}
          \and
          School of Control and Computer Engineering, North China Electric Power University, Beijing, 071003, China\label{addr2}
          \and
          Faculty of Electronic and Information Engineering, Xi’an Jiaotong University, Xi’an, 710049, China\label{addr3}
          \and
          School of Mechanical Engineering, Shanghai Jiao Tong University, Shanghai, 200240, China\label{addr4}
          \and
          College of Aritificial Intelligence and Automation, Hohai University,  Nanjing, 211100, China\label{addr5}
}

\date{}

\maketitle

\begin{abstract}
Under extreme operating conditions—characterized by high particle multiplicity and heavily overlapping shower energy deposits—classical particle flow algorithms encounter pronounced limitations in resolution, efficiency, and accuracy. To address this challenge, this paper proposes and systematically evaluates a deep learning reconstruction framework: For multichannel sparse features, we design a hybrid loss function combining weighted mean squared error with structural similarity index, effectively balancing pixel-level accuracy and structural fidelity. By integrating 3D convolutions, Squeeze-and-Excitation channel attention, and Offset self-attention modules into baseline convolutional neural networks, we enhance the model's capability to capture cross-modal spatiotemporal correlations and energy-displacement nonlinearities. Validated on custom-constructed simulation data and Pythia jet datasets, the framework's 90K-parameter lightweight variant approaches the performance of 5M-parameter baselines, while the 25M-parameter 3D model achieves state-of-the-art results in both interpolation and extrapolation tasks. Comprehensive experiments quantitatively evaluate component contributions and provide performance-parameter trade-off guidelines. All core code and data processing scripts are open-sourced on a GitHub repository to facilitate community reproducibility and extension.
\end{abstract}

\section{Introduction}
The complexity and variability of high-energy collision events generated by the Large Hadron Collider (LHC) experiment create an urgent need for high-performance event reconstruction algorithms \cite{beaudette2014cms}. The particle flow (PFlow) algorithm was developed in the ALEPH experiment and then optimized in the CMS experiment of the LHC \cite{collaboration1995performance, brient2004particle, ruan2014arbor, cms2017particle}. It can significantly improve jet energy resolution, missing transverse energy (MET) measurement, and physical object identification by effectively distinguishing between the energy deposition of neutral particles and charged particles in the calorimeter \cite{collaboration2009particle, collaboration2010commissioning, cms2011determination}. However, in extreme environments with high particle densities or overlapping energy deposits, the classical PFlow algorithm still has limitations. These challenges have prompted researchers to seek more advanced methods to improve the efficiency of energy reconstruction.

In recent years, deep learning techniques have reinvigorated PFlow research. Many recent works have transformed the quantum energizer reconstruction problem into a computer vision task by treating the detector's high-dimensional readout data as an image to be processed with the help of neural networks. For example, the study Towards a Computer Vision Particle Flow \cite{di2021towards} shows that the analysis of quantum energizer images based on convolutional neural networks (CNNs) \cite{cogan2015jet, de2016jet} can effectively improve the energy reconstruction of neutral particles, and maintains good performance even in the case of high overlap between the energy deposits of neutral and charged particles. Even in the case of high overlap of energy deposition between neutral and charged particles, the performance can be maintained. The CMS experiments also tried the graph neural network (GNN) approach to train a learnable whole-particle reconstruction model by fusing tracker and quantizer information, which outperformed the traditional PFlow algorithm \cite{di2021towards, qasim2019learning}. These studies demonstrate the powerful potential of deep learning in mining complex patterns of detector data, and provide new directions for breaking the bottleneck of traditional methods and expanding the scope of PFlow applications \cite{pata2023machine}.

In this study, we have further explore several innovative designs of network structures and training strategies based on the above research to enhance the performance of deep learning based energy reconstruction for quantum energizers. The training of our model is based on the simulation data set, and the movement of various particles in the experiment is simulated by code. Meanwhile, to evaluate the robustness of the model, we used the jet data generated by the Pythia event generator as the test set, which does not involve model training. This can test the generalization performance of the model under extreme conditions \cite{baldi2014searching, mokhtar2023progress, evans2008lhc}.

We propose a multimodal loss function to train the network model. Compared to using only the traditional mean square error (MSE) or weighted MSE, we combine it with the structural similarity index (SSIM) \cite{wang2004image} to form a multimodal loss function combination that focuses on both pixel-level error and image structural fidelity. SSIM originates from the field of image processing, and is used to measure the structural consistency between predicted images and real images. Experiments show that the multimodal loss incorporated into SSIM achieves a better balance between the overall energy accuracy and the quality of the structural reconstruction,  and exhibits high stability in the face of complex or infrequent input scenarios (e.g., jet injection).

On the basis of optimizing the loss function, we introduce a three-dimensional convolutional neural network module (3D CNN) \cite{ji20123d} to process the quantum energizer data as a three-dimensional image with a depth structure, so as to capture the spatial correlation features of the particle clusters shooting between multiple layers .Such 3D feature extraction approach has been proven to have strong performance in tasks such as particle cluster jet identification, and can significantly improve the model's understanding of the development process of particle energy deposition . Secondly, we introduce a channel attention mechanism, the Squeeze-and-Excitation (SE) module \cite{hu2018squeeze}, which adaptively strengthens the response of the most informative channel by modeling the dependencies between channels, and is widely recognized to improve the reconstruction quality and robustness of the model in image reconstruction tasks. Subsequently, we further introduce the row self-attention mechanism for the nonlinear offset phenomenon of high-energy particles. Since the particle offset distance is non-constant with energy, the self-attention mechanism \cite{vaswani2017attention} calculates the global correlation between the feature positions and establishes the energy-displacement mapping relationship under different offset scales, so as to accurately characterize the azimuthal offset characteristics of the particle trajectories.

To validate the effectiveness of the above design, we conducted systematic ablation experiments on the baseline CNN architecture \cite{meyes2019ablation}. The results show that each module individually improves the model performance and demonstrates improvement in generalization ability under highly sparse input conditions.

\section{Simulation experiments and datasets}
In view of the high requirements of complete Monte Carlo simulation in terms of computational resources and implementation complexity, this study adopts a modeling strategy based on a simplified stochastic process to generate energy distribution maps of the responses of various types of detectors of particles in high-energy collisions. The simulation method is based on the basic physical laws, reasonable sampling of particle types, charge states and their momentum distributions, and parameterized to simulate their energy deposition processes in the trajectory detector, electromagnetic quantum energizer (EMCal) and hadronic quantum energizer (HCal), supplemented by the deflection effect of charged particles under the action of a magnetic field, so as to construct a synthetic dataset that conforms to the physical characteristics. The final output data is in the form of a gray-scale image with a spatial resolution of 56×56, which not only retains the core physical characteristics of the particle detection response, but also has good computational efficiency, which is suitable for the application of data-driven methods such as deep learning in the reconstruction and analysis of particle physics images.

The various types of images in the dataset correspond to the responses of different detector systems in high-energy physics experiments: the Emcal image simulates the energy measurement process of electromagnetic quantum devices for particles such as electrons and photons; the Hcal image reflects the response of hadron quantum devices to the energy of protons, neutrons, and other hadrons. Tracker\_p and Tracker\_n record the trajectories of positive and negative charged particles in the magnetic field respectively, which are used to infer their momentum and charge; and the Truth image provides the real reference value of the particle energy distribution, which serves as an ideal benchmark for evaluating the performance of the model.

Two datasets with different physical characteristics were constructed in this study.
The Gauss\_S1.00\_NL0.30\_B0.50 dataset contains 10,000 image samples, each consisting of five types of images, namely, Emcal, Hcal, Tracker\_n, Tracker\_p, and Truth, which are used to simulate the energy response characteristics of various types of detectors in the background of standard particles.
The Gauss\_S1.00\_NL0.30\_B0.50\_jet dataset is extended from the former, containing 1,000 samples, and introducing Jet images generated by Pythia 8 on top of the original images to form a combination of six types of images: Emcal, Hcal, Tracker\_n, Tracker\_p, Jet and Truth. A combination of six types of images. It should be noted that the Jet dataset was used for testing purposes only and was not involved in the training process of the model.

\begin{figure*}
\centering
\includegraphics[width=\textwidth]{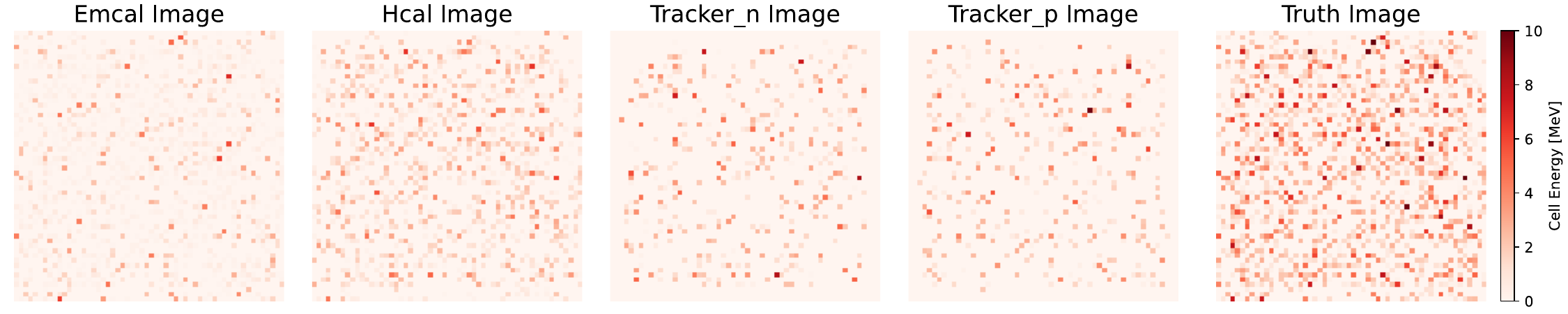}
\caption{Simulation data display, from left to right, they are respectively Emcal, Hcal, Tracker\_n, Tracker\_p, and Truth Image. The darkness of the pixel grid color represents the detected energy at the corresponding position.}
\label{example}
\end{figure*}

\textcolor{blue}{Figure~\ref{example}} is an example of one dataset.

\section{Design of Deep Neural Networks}
\subsection{Design of Deep Neural Networks}

\begin{figure*}
\centering
\includegraphics[width=\textwidth]{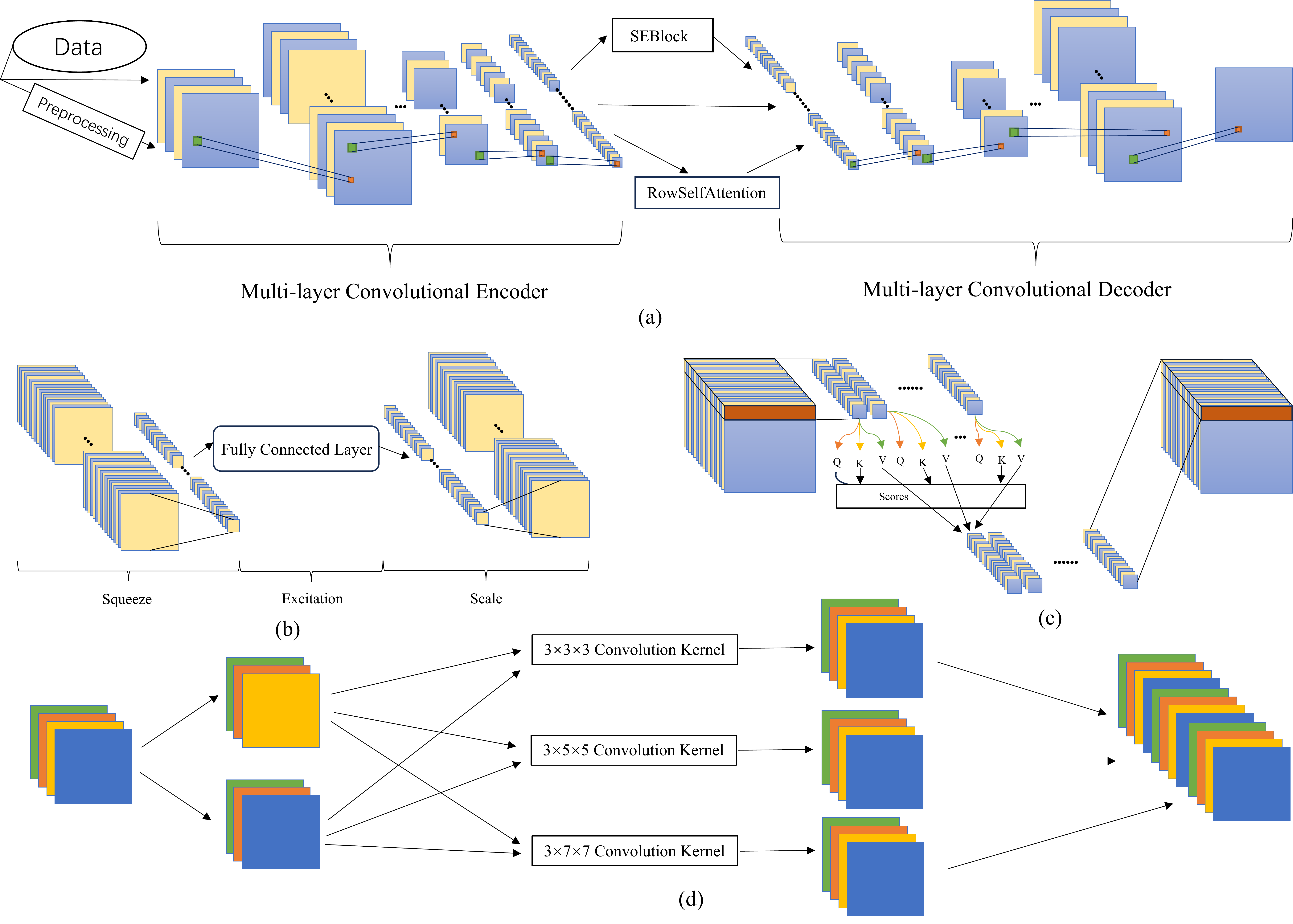}
\caption{}
\label{ms}
\end{figure*}

The structures of the models are shown in the \textcolor{blue}{Figure~\ref{ms}}.

\subsubsection{CNN}
The baseline CNN model adopts a standard encoder-decoder architecture: four convolutional layers form the encoder to extract higher-order features step by step, and three convolutional layers form the decoder to reduce the output energy distribution map step by step. The input is a 4-channel detector image containing electromagnetic calorimeter (Emcal), hadronic calorimeter (Hcal), negative particle tracker (Tracker\_n) and positive particle tracker (Tracker\_p), and the output is the reconstructed particle energy map. The convolutional neural network has the properties of local sensing and translation invariance, in which the local sensing field enables the model to effectively capture the local patterns of particle energy deposition, and the translation invariance brought by weight sharing ensures that the features extracted by the convolutional kernel are independent of the position. Therefore, even if the particle traces are shifted by a short distance on the detector image, the CNN maintains the robust modeling and reconstruction capability of the corresponding energy deposition patterns.

\subsubsection{Channel Attention Module}
A Squeeze-and-Excitation Channel Attention Block (SEBlock) is inserted between the encoder and decoder of the baseline CNN architecture. This module first performs global average pooling on the feature maps output from the encoder, compressing the 2D features of each channel into a scalar that characterizes the global semantics of the channel; and then generates the weights of each channel through a bottleneck network consisting of two fully connected layers. The output of the second fully-connected layer is compressed to the interval of 0~1 by sigmoid activation to obtain the importance coefficient of each channel, and then the original feature map is multiplied by the corresponding coefficient channel by channel to complete the feature weight calibration. By fusing the global semantic information to adaptively adjust the channel response, SEBlock is able to enhance the channel features containing important information and suppress redundant features. With high channel count calorimeter data input, the introduction of SEBlock can direct the model to focus on the detector channels that contribute more to the energy reconstruction, and improve the modeling differentiation of signals from different detector channels.

\subsubsection{3D Convolution-based Feature Extraction Module }
The model introduces a 3D convolution processing mechanism at the input stage. The positive and negative particle-related detector image channels are combined into two three-channel groups (each containing Emcal, Hcal, and Tracker\_n or Tracker\_p), respectively. Subsequently, 3D convolutional filtering (e.g., 3×3×3, 3×5×5, 3×7×7 convolution kernels) is applied to each three-channel group to simultaneously extract features in both spatial and cross-channel dimensions. 3D convolution utilizes the channel dimension as the “depth” for the convolution operation to capture the cross-modal correlation patterns within the channel group at one time, and then extracts the discriminative features with a temporal-like structure, which helps the model to understand the physical relationship between Emcal, Hcal and Tracker. The introduction of a multi-scale convolution kernel enables the model to capture the patterns of energy deposition of positive and negative particles at different spatial scales, resulting in a richer feature representation. In contrast, the traditional 2D convolution only extracts local features independently on each channel, and then linearly combines the channel signals, which may lead to confusion of information from different detector channels. By introducing 3D convolution into the positive and negative particle branches and explicitly modeling the channel dimensions, the model effectively eliminates the interference caused by 2D convolution channel mixing, and greatly improves the ability to characterize the difference in energy deposition of positive and negative particles. This 3D convolution mechanism significantly enhances the model's ability to capture and reconstruct complex patterns in multi-channel input data.

\subsubsection{Offset Self-Attention Module}
To address the phenomenon of non-constant-length offsets of energetic particles in the offset direction, we propose the offset self-attention mechanism \cite{vaswani2017attention}. This module dynamically captures the influence of the energy of the particles with respect to the offset pattern by globally modeling the long-range dependencies among the features in the rows. Specifically, the input feature map is first reorganized into multiple sequences along the main offset direction of the particle, and then the energy-displacement mapping relationship is adaptively established by self-attention to compute the global correlation between the positions within the rows. The local receptive fields of traditional CNNs are difficult to capture the non-constant-length shift of the particle offset distance due to the change of particle energy (the offset distance is inversely proportional to the energy). Row attention directly models the association strength of any two positions within a row through global autocorrelation computation, breaking through the size limit of the convolutional kernel. Attention weights dynamically reflect the nonlinear relationship between particle energy and offset distance: high-energy particles with short offset distances are modeled by focusing on neighboring regions through strong localized attention; low-energy particles with long offset distances are modeled by dispersing their attention to cover a wider range, realizing energy-adaptive offset compensation.

\subsection{Loss Function Design}
The loss function is used to quantify the difference between the predictions of the model and the objective as a goal to optimize the algorithm during training. The choice of loss function has a non-negligible impact on the model effect. Exploring a better loss function than is one of the core purposes of the experimental investigation in this paper.

Ordinary mean square error is often used in image reconstruction tasks to measure the pixel-level error between the predicted and true values. It calculates the average of the squared errors between the predicted and true values and can provide a smooth optimization objective. The formula is shown in \textcolor{blue}{Equation~\ref{MSE}}.where $N$ is the total number of pixel sampling points in the prediction map, $y_{pred,i}$ and  $y_{true,i}$, are the pixel values output by the model and the target pixel values.

\begin{equation}
\mathcal{L}_{\text{MSE}} = \frac{1}{N} \sum_{i=1}^{N} \left( y_{\text{pred},i} - y_{\text{true},i} \right)^2 
\label{MSE}
\end{equation}

Considering that the essence of the energy reconstruction task is the processing of multi-channel sparse data, the ordinary mean-square error will inevitably be affected by more than half of the zeros in the prediction, which will lead to a limited accuracy of the energy reconstruction of the model. The weighted mean square error is designed to completely eliminate the influence of zero values and is able to pay higher attention to pixel points with larger true values, thus achieving more efficient exploitation of the model's potential. The formula is shown in \textcolor{blue}{Equation~\ref{LSME}}, where $\epsilon$  is the stabilization factor, usually set to $10^{-8}$ to prevent the denominator from being zero.

\begin{equation}
\mathcal{L}_{\text{WMSE}} = \frac{\sum_{i=1}^{N} y_{\text{true},i} (y_{\text{pred},i} - y_{\text{true},i})^2}{\sum_{i=1}^{N} y_{\text{true},i} + \epsilon}
\label{LSME}
\end{equation}

SSIM is a measure of structural similarity between two images, preferring to evaluate the quality of the generated image in terms of brightness, contrast and structural information. $\mu_{\text{pred}}$, $\mu_{\text{true}}$ are the mean values of the predicted and real images respectively, which are used to measure the overall brightness. $\sigma_{\text{pred}}^2$, $\sigma_{\text{true}}^2$ are the variance of the predicted image and the real image respectively, which is used to measure the contrast of the image. $\sigma_{\text{pred,true}}$ is the covariance between the predicted image and the real image, which is used to measure the similarity of the structural information of the image. $C_1$, $C_2$ are the stabilizing factors, which are used to prevent the denominator from going to zero, and they are usually set as: $C_1 = (0.01L)^2, C_2 = (0.03L)^2$, where $L$ denotes the dynamic range of the pixel value. Finally, SSIM is calculated as \textcolor{blue}{Equation~\ref{SSIM}}.

\begin{equation}
\text{SSIM}(y_{\text{pred}}, y_{\text{true}}) = \frac{(2\mu_{\text{pred}}\mu_{\text{true}} + C_1)(2\sigma_{\text{pred},\text{true}} + C_2)}{(\mu_{\text{pred}}^2 + \mu_{\text{true}}^2 + C_1)(\sigma_{\text{pred}}^2 + \sigma_{\text{true}}^2 + C_2)}
\label{SSIM}
\end{equation}

The closer the SSIM index is to 1, the more similar the image structures are, so $\mathcal{L}_{\text{SSIM}}$ is defined as \textcolor{blue}{Equation~\ref{LSSIM}}.

\begin{equation}
\mathcal{L}_{\text{SSIM}} = 1 - \text{SSIM}(y_{\text{pred}}, y_{\text{true}})
\label{LSSIM}
\end{equation}

\section{Experimental Preparation}
\subsection{Training Strategies}
In the design of the experimental training strategy, we first preprocess the input data uniformly. Specifically, for the four types of input images (Tracker\_n, Tracker\_p, Emcal, and Hcal), min-max normalization is implemented over the entire range of the dataset to ensure that the energy response values of the different modal detector channels are modeled at the same numerical scale, so as to improve the model's convergence stability and feature fusion capability.

During the training process, we adopt the optimization strategy of gradually adjusting the learning rate combined with the early stopping mechanism. The initial learning rate is set to 0.001, and automatically decays to half of the original rate after every 30 training epochs to improve the convergence accuracy of the model in the later training phase. In the formal training phase, the early stopping criterion is set to terminate the training if the validation loss does not improve within 20 consecutive epochs to ensure that the model is adequately trained.

All experiments were performed on a single NVIDIA RTX 3090 GPU (24GB of graphics memory), ensuring effective support for high-dimensional image inputs and high-capacity model structures.

\subsection{Model Evaluation Indicators}
\subsubsection{General Energy Reconstruction Task Evaluation Indicators}
\textbf{\textit{Weighted Mean Square Error (WMSE)}}, which is used to reflect differences in the importance of the true energy of different samples in the overall assessment. This metric focuses the assessment more on the prediction accuracy of key samples by giving higher energy samples more weight, with smaller values in absolute terms being better.

\textbf{\textit{Relative Energy Residual (RER):}} measures the percentage deviation of the predicted total energy relative to the true total energy. A positive value of this indicator indicates that the model overestimates the energy, and a negative value indicates that the model underestimates the energy, with smaller absolute values being better.

\textbf{\textit{Euclidean Distance Error between Image Centers of Mass (Centroid Distance Error): }}Images of predicted and true energy distributions are treated as two-dimensional energy distributions. The position of the energy center of mass is calculated on each image, and then the Euclidean distance between the predicted center of mass and the true center of mass is calculated. This distance error will be larger if the spatial distribution of energy in the predicted image deviates from the true situation.

\subsubsection{Jet Energy Reconstruction Assessment Metrics and Process}

The jet energy reconstruction task involves extracting particle momentum and performing jet clustering at the image level, and then evaluating the reconstruction errors in jet energy and position. The whole process can be summarized in the following stages:\textit{Image pixels → particle momentum calculation → FastJet clustering analysis → jet mapping and matching → metrics calculation. }

The details of each stage are designed as follows:

\textbf{Particle Momentum Calculation:} Each pixel of the 56$\times$56 energy image is considered to be a particle. The energy value of each pixel corresponds to the energy of the particle, $E$. Assuming that the particles are all from colliding vertices, the 3D momentum vector and position coordinates of the particles are derived from the pixel positions.For each pixel particle, we denote its four-momentum as $(E, p_x, p_y, p_z)$ and position as $(x, y, z)$, Where $p_x, p_y, p_z$ can be computed from the particle energy E and known geometric relations. This step converts the static energy distribution into a kinematically tractable list of particles.

\textbf{FastJet Cluster Analysis (anti-kt algorithm): }The particle list information is entered into the FastJet software, and the particles are reconstructed into a jet using the anti-kt clustering algorithm, which combines the particles to form a jet based on the interparticle distance and momentum. Through the FastJet cluster analysis, we obtain a number of reconstructed jets, each of which has the parameters of fast velocity $y_i$ (pseudo-rapidity) and azimuth $\varphi_i$, transverse momentum $P_{T,i}$, energy $E_i$, and area $A_i$ (which represents the size of the area covered by the jet in the $(y,\varphi)$ space), which describe the kinematic properties of the jet and its extent in the $(y,\varphi)$ plane. 

\textbf{Jet Mapping and Matching:} In order to visualize the correspondence between the jet and the original image, we map each reconstructed jet as a 2D circular area equal to the area computed by FastJet as $A_i$. The radius of the area of the circle is defined as r,  such that the area of the circle $\pi r^2 = A_i$, i.e., $r=\sqrt{A_i/\pi}$. In this way, each jet can be approximated as a circle in the image plane (corresponding to its capture range in $(y,\varphi)$ space). Next, the reconstructed jets need to be matched with real jets: by calculating the overlap area of the circular region, we pair the predicted Jet with the closest real jet. The matching criterion is usually to maximize the overlap area or to use the correspondence with the smallest $\Delta R$  (distance measure). After matching, we can consider each pair of jets (predicted vs. true) to represent a reconstruction of the same physical injection.

\textbf{Jet Evaluation Metrics Calculation:} For each matched jet pair, we further calculate the following two evaluation metrics:

\textbf{Relative Jet Energy Error:} A measure of the accuracy of the jet energy reconstruction, defined as the relative proportion of the deviation of the predicted jet energy from the true jet energy, as \textcolor{blue}{Equation~\ref{jet error}}, where $E_{i_{\mathrm{predict}}^{*}} $ and $E_{i_{\mathrm{truth}}^{*}}$ are the predicted and true values of the \(i\)-th Jet energy, respectively.

\begin{equation}
    \text{Relative Energy Error} = \frac{E_{i_{\mathrm{predict}}^{*}}-E_{i_{\mathrm{truth}}^{*}}}{E_{i_{\mathrm{truth}}^{*}}} \times 100\%
\label{jet error}
\end{equation}

The closer the value is to 0, the better, with positive values indicating high predicted energy and negative values indicating low energy.

\textbf{Jet Spatial Position Difference:} A measure of the reconstruction error of the Jet's position in three-dimensional space. The coordinates (x, y, z) of the position in space are first deduced from the matched jet parameters. The orientation of the jet axis in the detector coordinate system is computed by combining the rapidity y with the azimuthal angle $\varphi$, and then extended to the energy-dependent spatial position. The Euclidean distance between the predicted and true Jet positions is then calculated as \textcolor{blue}{Equation~\ref{distance}}.

\begin{equation}  \text{Position Difference}=\sqrt{(x_{i^*_{\text{truth}}} - x_{i^*_{\text{predict}}})^2 + (y_{i^*_{\text{truth}}} - y_{i^*_{\text{predict}}})^2 + (z_{i^*_{\text{truth}}} - z_{i^*_{\text{predict}}})^2}
\label{distance}
\end{equation}

This distance measures how far the Jet axis (or center of mass) has shifted in space, with smaller distances indicating a more accurate positional reconstruction.

\section{Experiment}
\subsection{Optimal loss function exploration}

This experiment evaluates the effect of different loss functions on the performance of the CNN baseline model, and MSE, WMSE, SSIM and its hybrid form (MIX: WMSE+$\lambda$SSIM, $\lambda$=0.05) are selected for comparative analysis. The models are validated for their localization accuracy and energy prediction robustness on ordinary datasets and test sets containing extreme physical phenomena (jet events), respectively. The experimental results are shown in \textcolor{blue}{Figure~\ref{2}} and \textcolor{blue}{Figure~\ref{3}}.

\begin{figure*}
    \centering
    \begin{minipage}{0.5\textwidth}
        \centering
        \includegraphics[width=\textwidth]{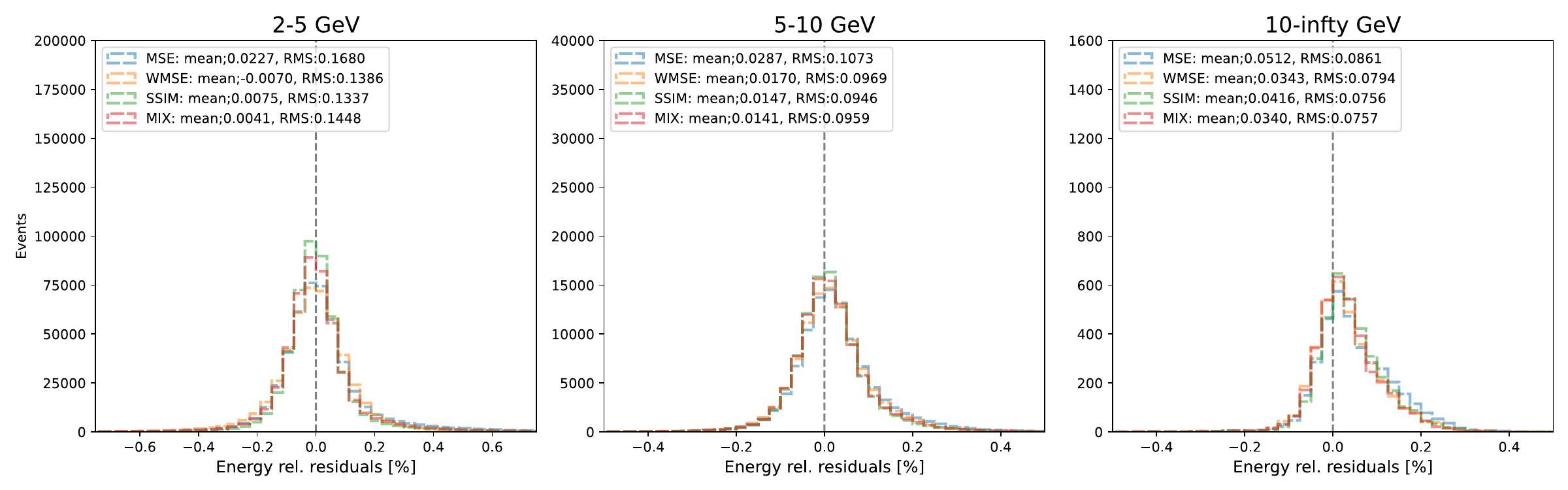}
    \end{minipage}%
    \begin{minipage}{0.5\textwidth}
        \centering
        \includegraphics[width=\textwidth]{energy-bset-loss.pdf}
    \end{minipage}
    \caption{}
    \label{3}
\end{figure*}

\begin{figure*}
    \centering
    \begin{minipage}{0.5\textwidth}
        \centering
        \includegraphics[width=\textwidth]{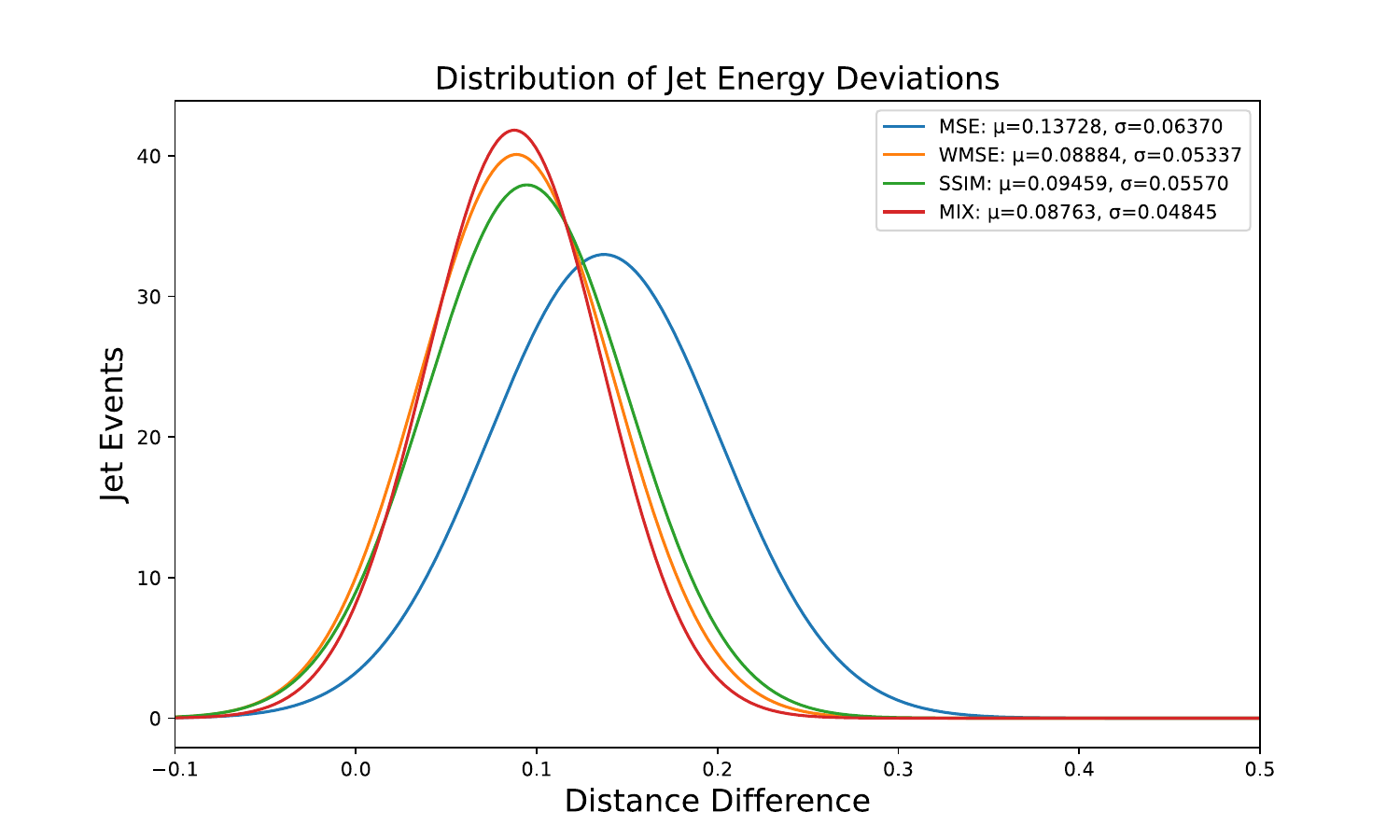}
    \end{minipage}%
    \begin{minipage}{0.5\textwidth}
        \centering
        \includegraphics[width=\textwidth]{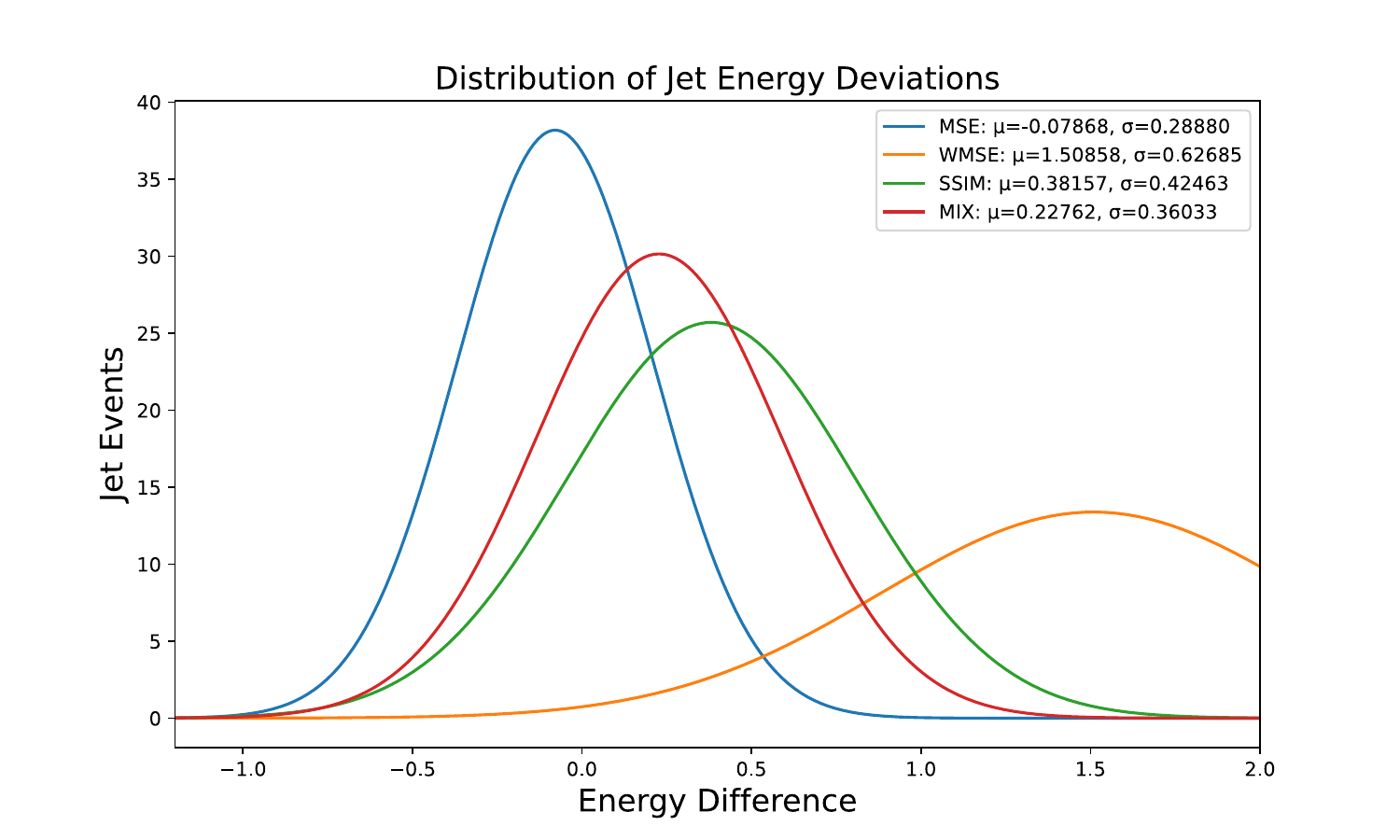}
    \end{minipage}
    \caption{}
    \label{3}
\end{figure*}

The model using the MSE loss function exhibits relatively high errors in the common interpolation task. In terms of relative energy residuals, the mean and STD of the MSE model are higher than those of the other loss functions, indicating that its reconstruction of the energy values in the high-energy region is less accurate.The MSE model also has the largest inter-center-of-mass Euclidean distance error, reflecting the large bias of MSE on the reconstruction of the overall information. The reason for this phenomenon is that MSE tends to minimize the global squared error and pays too much attention to the zero values in the background, which affects the model's holistic relational learning of the energy in the non-zero region. In the extreme jet extrapolation task, which occurs in the test set only, the MSE model struggles to accurately capture new high-energy jets, which manifests itself in the largest jet center-of-mass spatial distance error. At the same time, the MSE's jet energy relative error instead remains low (showing a slight underestimation) due to a more conservative prediction of unknown high-energy values, but this low error comes at the cost of failing to reconstruct the full energy of the jet.

The introduction of weighted MSE losses significantly improves the reconstruction performance of the model in the interpolation task. Compared with MSE, the WMSE model has smaller relative energy residuals and significantly lower inter-centre-of-mass distance errors, suggesting that giving higher weights to non-zero high-energy pixels can guide the model to focus more on the pixel information in the high-energy region and mitigate the dominant role of the zero background on the loss. However, in the extrapolation task containing extreme jets, WMSE shows excessive sensitivity to extreme sparse data: while the model continues to accurately locate the jet position (the jet centre-of-mass spatial distance error remains small), its estimation of the jet energy is severely biased. Its jet energy relative error is much higher than the other loss functions, reflecting the fact that overemphasising sparse high-energy points leads to unstable model predictions under unknown distributions and significant overestimation of jet energy.

The model with SSIM structural similarity loss achieves higher reconstruction accuracy in the interpolation task. The relative energy residuals are significantly reduced and the total energy estimation is more accurate. The center-of-mass localization error is also smaller and close to the optimal level. The SSIM loss focuses on the local structural information of the image and avoids the limitation of relying on the optimization of point-by-point intensity error only, and thus has higher robustness to sparse data, and is less likely to overlook details due to the majority of background zeros.In the extrapolation scenario of extreme jet, the SSIM model is still able to identify and reconstruct the jet structure more accurately, with the jet center-of-mass spatial distance error close to the level of WMSE. However, its prediction of the jet energy is slightly overpredicted, with a certain degree of relative error in the jet energy, but compared to the serious misalignment of WMSE, it reflects the relatively robust reconstruction ability of SSIM in unknown sparse scenarios.

MIX model combines the two objectives of MSE and SSIM, taking into account both global error minimization and local structure fidelity, and demonstrates comprehensively optimal performance in the interpolation task. The relative energy residuals of this model are almost comparable to those of the SSIM model, while the inter-center-of-mass Euclidean distance error is the lowest among all models, achieving simultaneous improvements in energy values and global structure reconstruction. Due to the advantages of incorporating multiple losses, the MIX model also shows greater adaptability in the extrapolation task: for unknown extreme jets, MIX almost completely maintains the Jet center-of-mass spatial distance error of WMSE, and accurately captures the positional distribution of the jets; at the same time, its jet energy prediction has a small bias, and the Jet energy relative error is significantly lower than that of the SSIM and WMSE models, with no serious The relative error of Jet energy is significantly lower than that of SSIM and WMSE models, without serious overestimation or underestimation. On the whole, the hybrid loss strategy effectively balances the various indicators and shows robust optimization advantages in complex scenarios.

In summary, various loss functions have their own advantages and disadvantages in the energy reconstruction task: MSE is computationally stable as a baseline method, but it is difficult to accurately reconstruct details in the higher energy region due to the bias towards the background zeros; WMSE strengthens the focus on the high-energy particle information, which improves the reconstruction accuracy under regular data, but is overly sensitive to the distributional variations, which may lead to poor consistency under extreme sparsity; SSIM focuses on the structure of the image, which makes the energy and shape reconstruction more accurate and robust to data sparsity, but still suffers from a slight bias under unseen extreme injection; MIX combines the advantages of multiple objectives and achieves optimal or near-optimal results in both energy and position, realizing a comprehensive and balanced reconstruction performance. Future work may consider introducing adaptive adjustment of loss weights (e.g., dynamically adjusting the $\lambda$ coefficient in the hybrid loss), which automatically balances the roles of each loss term according to the training stage or data distribution. In addition, exploring reconstruction strategies that fuse multimodal information is also a promising avenue, such as combining other measurements of the injection with image information to further improve the model's reconstruction accuracy of complex injection patterns and generalization ability to unknown scenarios.

\subsection{Optimal Module Exploration}
This experiment explores the improvement of the baseline model performance of the three modules under different parametric quantities, which leads to the optimal model selection scheme. It should be further clarified that the improvement of this experiment is reflected in two aspects, i.e. the improvement of the inference performance and the improvement of the extrapolation performance.

\subsubsection{Improvements in inference performance}
In this module, we select relative energy residuals, Euclidean distance error between image centers of mass and weighted mean square error as evaluation metrics to quantify the model's localization accuracy and energy prediction performance on common datasets. In addition, in order to evaluate the model performance in more detail, we will evaluate the performance of the model on three energy intervals, namely, low (2-5 GeV), medium (5-10 GeV), and high (above 10 GeV), respectively.

\begin{figure*}
\centering
\includegraphics[width=\textwidth]{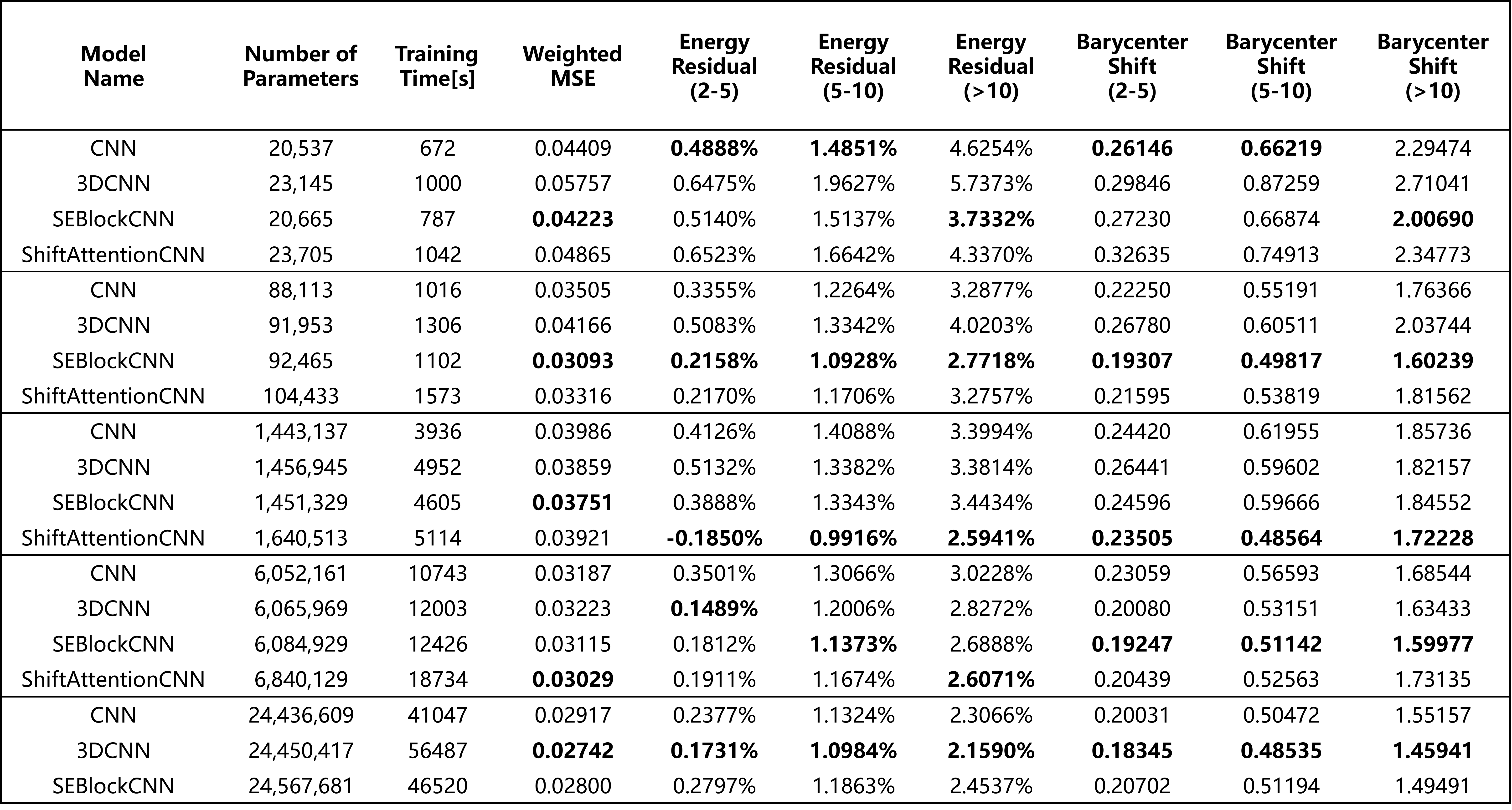}
\caption{}
\label{tb2}
\end{figure*}

The experimental results are shown in \textcolor{blue}{Figure~\ref{tb2}}.

From the experimental results, it can be seen that SEBlockCNN has the best overall performance in all interpolation metrics at about 20,000 parameters, with particularly good localization accuracy and energy prediction performance in the high-energy section. the improvement of ShiftAttentionCNN is confined to the high-energy section, whereas the 3DCNN is inferior to the Baseline model in most of the metrics.

With 90,000 parameters, SEBlockCNN outperforms in all the metrics and achieves the maximum performance improvement with more than 10\% improvement in all the metrics; ShiftAttentionCNN also has a moderate improvement, but the improvement is smaller; and 3DCNN is still inferior to Baseline.

For a model with about 1 million parameters, all three improved models show slight improvements in the metric of WMSE. In terms of energy residuals, ShiftAttentionCNN has the most significant effect, with a reduction of about 30\% in the medium-energy band, a reduction of about 24\% in the high-energy band, and an improvement in residuals in the low-energy band that is close to zero. SEBlockCNN and 3DCNN show essentially no improvement in this area. The center-of-mass shift results are similar to the energy residuals. Taken together, ShiftAttentionCNN shows the most significant improvement in the interpolation performance, while SEBlockCNN and 3DCNN only slightly improve in some metrics.

At a parameter scale of about 5 million, the WMSE of all three improved models improved or equalized with baseline. In terms of energy residuals in each energy band, all improved models outperform Baseline, with the most significant residuals in the low-energy band: the 3DCNN improvement comes to 57\%. For the medium energy band Residual, all the three improved models have an improvement of around 10\%. The high-energy Residual also shows an improvement of 7-14\%. The center-of-mass offset metric shows a similar trend. Taken together, all the improved models outperform Baseline in most of the metrics, with ShiftAttentionCNN outperforming in key metrics such as Overall Error and High Energy Residual; SEBlockCNN shows great improvement in the center-of-mass offset metrics in the low and middle energy bands; and 3DCNN also achieves overall performance improvement, but its improvement is slightly smaller than that of the other two models. 3DCNN also achieves a comprehensive performance improvement, but its improvement is slightly smaller than that of the other two models.

With about 25 million parameters, 3DCNN has the best inference performance and outperforms Baseline in all metrics, while SEBlockCNN saturates the improvement and is comparable to Baseline or slightly degraded in most of the metrics, with only a slight advantage in individual metrics such as weighted MSE.

\subsubsection{Extrapolation performance improvement}
This experiment evaluates the impact of three modules on the performance of the baseline model under different parametric quantities. For this module, we select Jet energy relative error and Jet spatial position difference as evaluation metrics to quantify the model's localization accuracy and energy prediction performance on datasets containing extreme physical phenomena (jet events).

\begin{figure*}
\centering
\includegraphics[width=\textwidth]{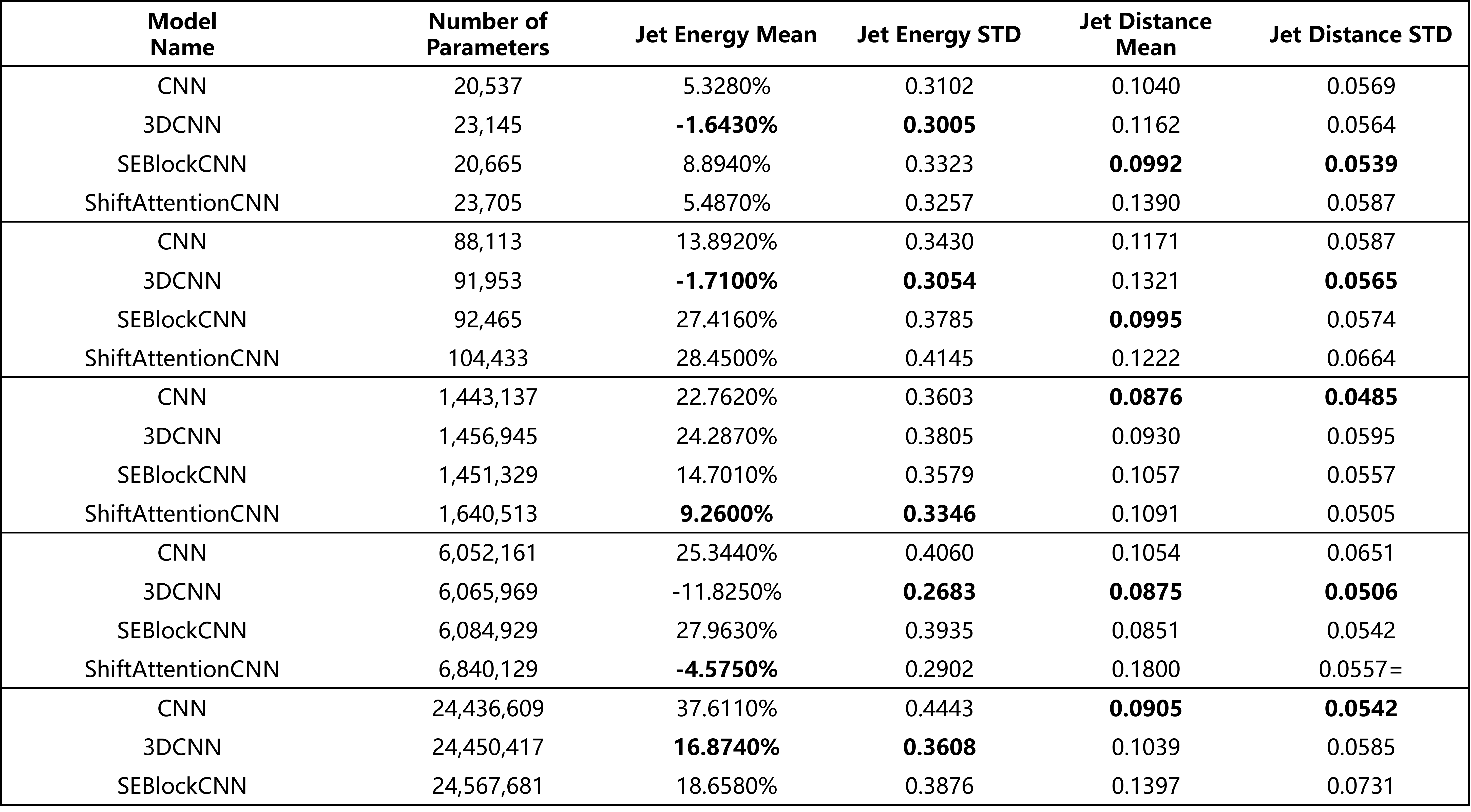}
\caption{}
\label{tb3}
\end{figure*}

The experimental results are shown in \textcolor{blue}{Figure~\ref{tb3}}.

For models with about 20,000 parameters, 3DCNN significantly reduces the critical Jet Energy Mean deviation, SEBlockCNN only improves on the Jet distance error without optimizing the critical energy deviation, and ShiftAttentionCNN does not improve on any of the metrics.

When the number of parameters increases to about 90,000, only 3DCNN shows an overall generalization advantage by significantly reducing the Jet Energy bias, while SEBlockCNN improves the spatial accuracy but cannot be considered as an overall enhancement due to the worsening energy bias, and ShiftAttentionCNN does not show any improvement in both energy and distance.

For the medium-sized model with 1 million parameters, ShiftAttentionCNN and SEBlockCNN achieve an overall improvement in energy prediction at the cost of a slight increase in spatial error, with ShiftAttentionCNN showing the greatest improvement in Jet Energy; conversely, 3DCNN fails to improve energy generalization and slightly reduces the spatial accuracy, which is not advantageous at this parameter scale.

At about 5 million parameters, ShiftAttentionCNN greatly improves the energy bias but slightly increases the spatial error, 3DCNN improves both energy and distance, while SEBlockCNN fails to reduce the Jet energy bias in spite of improving the spatial prediction accuracy, and thus does not improve the basline CNN.

At about 25 million parameters, SEBlockCNN and 3DCNN still demonstrate better overall generalization performance under very large models: they significantly reduce the systematic bias in the injection energy and improve the reliability of the models in interpolation scenarios.

\begin{figure*}
    \centering
    \begin{minipage}{0.5\textwidth}
        \centering
        \includegraphics[width=\textwidth]{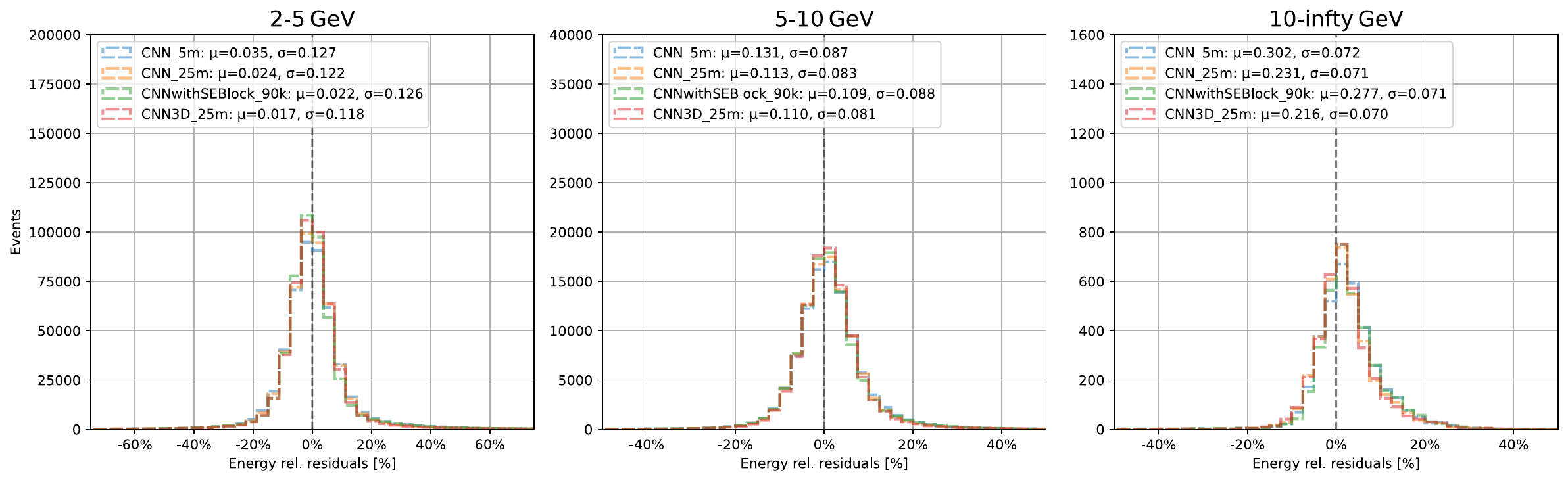}
    \end{minipage}%
    \begin{minipage}{0.5\textwidth}
        \centering
        \includegraphics[width=\textwidth]{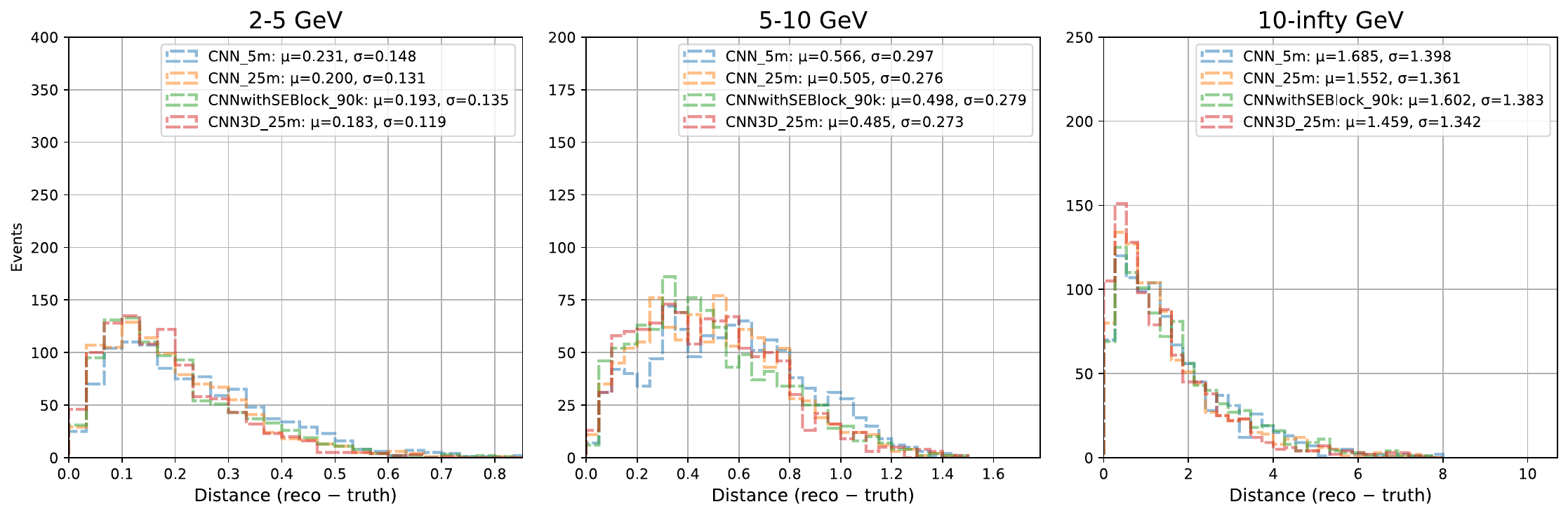}
    \end{minipage}
    \caption{}
    \label{4}
\end{figure*}

\begin{figure*}
    \centering
    \begin{minipage}{0.5\textwidth}
        \centering
        \includegraphics[width=\textwidth]{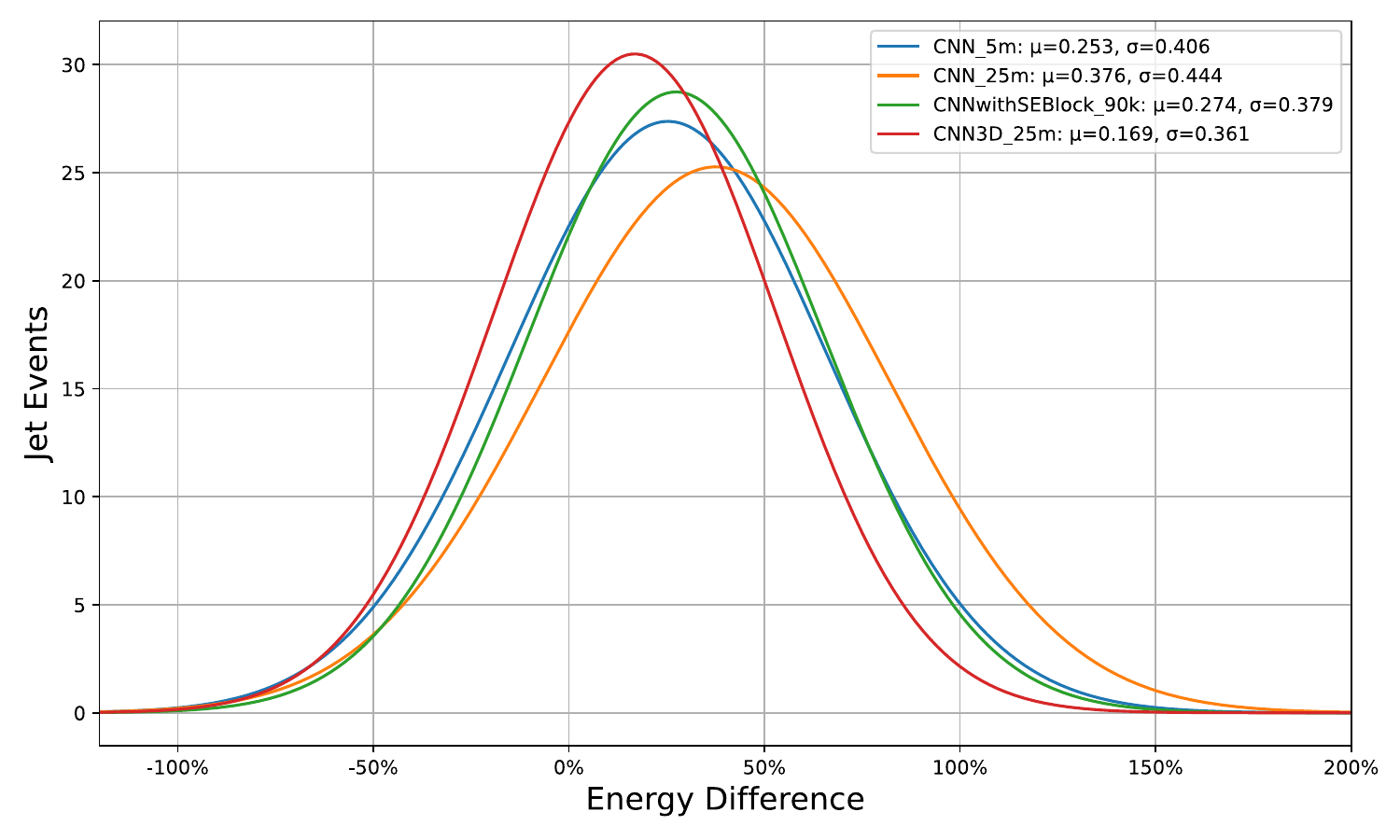}
    \end{minipage}%
    \begin{minipage}{0.5\textwidth}
        \centering
        \includegraphics[width=\textwidth]{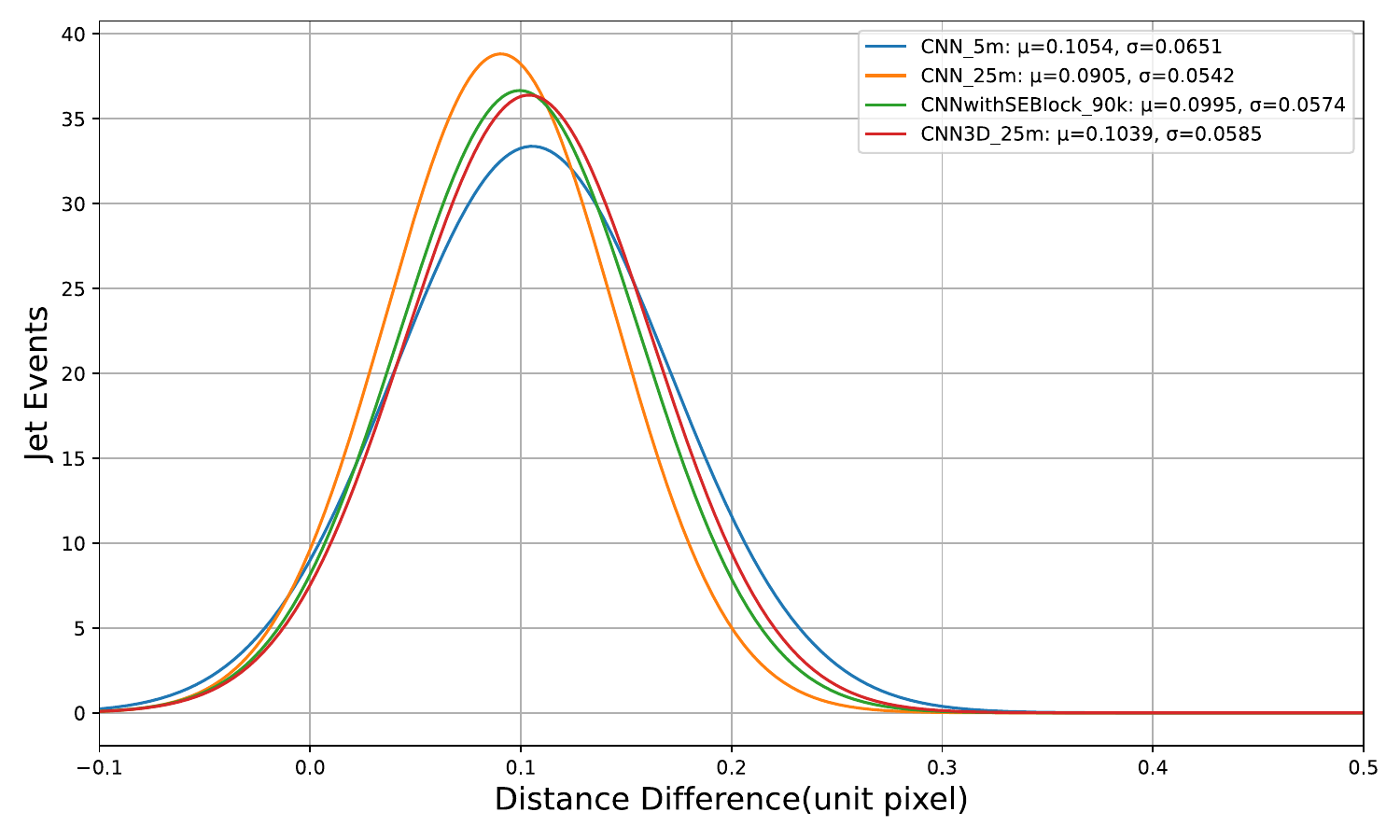}
    \end{minipage}
    \caption{}
    \label{5}
\end{figure*}

\subsection{Module validity argumentation}
All three improved modules outperform the baseline model in different scenarios.SEBlockCNN is suitable for small- to medium-scale inference tasks, especially in small- to medium-sized parameter counts (20,000 to 5,000,000), and its optimized feature weight distribution through the channel attention mechanism shows systematic improvement in energy residuals and center-of-mass offset metrics in low- and medium-energy bands. At very large parameter counts, the improvement tends to be saturated but not weaker than the baseline model. 3DCNN excels in its extrapolation performance by modeling the joint space-energy distribution through a three-dimensional convolution kernel, which strengthens the model's ability to generalize to extreme physical phenomena. Its inference performance also improves significantly when the number of parameters exceeds 1 million, indicating that 3D modeling requires sufficient parameter capacity to balance computational overhead and feature representation. ShiftAttentionCNN, on the other hand, performs best in high-energy segment inference and medium-scale extrapolation. It effectively captures long-range dependencies through the combination of shift operation and self-attention, e.g., at 1 million parameters, the energy residuals in the middle and high-energy bands are reduced by 30\% and 24\%, respectively. In addition, when the parameter capacity is sufficient (one million and above), ShiftAttentionCNN can also outperform the baseline model in terms of inference capability and extrapolation.

\subsection{Optimal model selection}
For the baseline model selection strategy, we constructed a dual benchmark system: CNNs with parameter sizes of 25,000,000 and 5,000,000 respectively were used as reference benchmarks at the same time.The 5M parameter model showed a clear inflection point of marginal benefit on the WMSE-parameter curve - 3.19 × 10\^-2→ 2.92×10\^-2 error improvement requires a 5-fold parameter growth, while the 25M model represents the theoretical performance limit under the current architecture, but exhibits severe overfitting during extrapolation tests. The two correspond to the optimal model capacity and the theoretical optimal reference frame, respectively. The establishment of this “ideal upper benchmark” makes it necessary for the new architecture to satisfy the cost-performance equilibrium point where (a) the accuracy completely exceeds the theoretical limit of the 25M benchmark model; and (b) the performance is not weaker than the theoretical limit of the 25M benchmark model, and significantly better than the 5M model in terms of the overall indexes. By setting up this evaluation index, it can effectively distinguish between real structural innovation and mere parameter stacking.

We propose two optimal improved models, one is the optimal model under full parameterization, the 25,000,000-parameter 3DCNN model, and the optimal model under restricted parameterization, the 90,000-parameter SEBlockCNN. SEBlockCNN. From the comparison results, it can be seen that 3DCNN-25M, with the advantage of 3D modeling, has stronger generalization ability while fully exceeding the theoretical limit of the 25M baseline model, which is especially suitable for high-precision and high-complexity tasks; and SEBlockCNN-90K, with a very low number of parameters to achieve the performance close to the theoretical optimum of the baseline model, proposes a light-weight solution.

The experimental results are shown in \textcolor{blue}{Figure~\ref{4}} and \textcolor{blue}{Figure~\ref{5}}.

\section{Conclusion}
This study systematically explores the impact of loss function design and model architecture innovation on the performance of high-energy physics image reconstruction tasks, revealing the key optimization paths and performance trade-off laws. Regarding the loss function, the hybrid loss function (WMSE+SSIM) demonstrates the best balance across multiple tasks, while the SSIM loss excels in structural fidelity. Regarding the innovative benefits of model architectures, channel attention plays extremely well in lightweight models, verifying the effectiveness of adaptive weighting of feature channels; 3D convolutional networks demonstrate unique balanced improvements in inference and extrapolation tasks, proving the generalization ability of 3D feature modeling to unknown physical scenes; and the Shift-Attention mechanism leads to a significant increase in reconstruction accuracy for high-energy segments. In addition, we establish the optimal configuration for this task: the 25M-parameter 3DCNN achieves state-of-the-art accuracy, while the 90K-parameter SEBlock model achieves nearly equivalent baseline model performance with 1/300 parameter count.

\bibliographystyle{cas-model2-names}


\end{document}